\documentclass[preprint,superscriptaddress,nofootinbib]{revtex4}
  
  \usepackage{amssymb,amsmath}
  \usepackage{graphicx}
  \usepackage{color}
  
\begin{document}
  \title{Possibility of searching for $B_{c}^{\ast}$ ${\to}$
        $B_{u,d,s}V$, $B_{u,d,s}P$ decays}
  \author{Junfeng Sun}
  \affiliation{Institute of Particle and Nuclear Physics,
              Henan Normal University, Xinxiang 453007, China}
  \author{Yueling Yang}
  \affiliation{Institute of Particle and Nuclear Physics,
              Henan Normal University, Xinxiang 453007, China}
  \author{Na Wang}
  \affiliation{Institute of Particle and Nuclear Physics,
              Henan Normal University, Xinxiang 453007, China}
  \author{Qin Chang}
  \affiliation{Institute of Particle and Nuclear Physics,
              Henan Normal University, Xinxiang 453007, China}
  \author{Gongru Lu}
  \affiliation{Institute of Particle and Nuclear Physics,
              Henan Normal University, Xinxiang 453007, China}
  \begin{abstract}
  The $B_{c}^{\ast}$ ${\to}$ $B_{u,d,s}V$, $B_{u,d,s}P$ decays are
  investigated with the QCD factorization approach, where
  $V$ and $P$ denote the ground $SU(3)$ vector and pseudoscalar mesons, respectively.
  The $B_{c}^{\ast}$ ${\to}$ $B_{u,d,s}$ transition form factors are
  calculated with the Wirbel-Stech-Bauer model.
  It is found that branching ratios for the color-favored and Cabibbo-favored
  $B_{c}^{\ast}$ ${\to}$ $B_{s}{\rho}$, $B_{s}{\pi}$ decays can reach up
  to ${\cal O}(10^{-7})$, which might be measurable in the future LHC
  experiments.
  \end{abstract}
   \maketitle

  \section{Introduction}
  \label{sec01}
  The vector $B_{c}^{{\ast}}$ meson, a spin-triplet ground state,
  consists of two heavy quarks with different flavor numbers
  $B$ $=$ $C$ $=$ ${\pm}1$, i.e., $\bar{b}c$ for $B_{c}^{{\ast}+}$
  meson and $b\bar{c}$ for $B_{c}^{{\ast}-}$ meson.
  With nonzero bottom and charm numbers, the bottom and charm
  quarks of the $B_{c}^{\ast}$ meson cannot annihilate
  into gluons and photons via the strong and electromagnetic
  interactions, respectively, unlike the decay
  modes of the unflavored $J/{\psi}(1S)$ and ${\Upsilon}(1S)$ mesons.
  The $B_{c}^{\ast}$ meson serves as a unique object in studying
  the heavy quark dynamics, which is inaccessible through both
  charmonium and bottomonium.

  The $B_{c}^{\ast}$ meson lies below the $B_{q}D_{q}$ ($q$ $=$ $u$, $d$, $s$)
  meson pair threshold.
  And the mass splitting $m_{B_{c}^{\ast}}$ $-$ $m_{B_{c}}$ ${\approx}$ 50 MeV
  \cite{prd86.094510} is less than the pion mass. Hence, the $B_{c}^{\ast}$
  meson decays via the strong interaction are strictly forbidden.
  The electromagnetic transition process, $B_{c}^{\ast}$ ${\to}$
  $B_{c}{\gamma}$, dominates the $B_{c}^{\ast}$ meson decays,
  but suffers seriously from a compact phase space suppression,
  which results in a lifetime of ${\tau}_{B_{c}^{\ast}}$
  ${\sim}$ ${\cal O}(10^{-17}\,{\rm s})$ \cite{epja52.90}.
  Besides, the $B_{c}^{\ast}$ meson decays via the weak interaction,
  although with very small decay rates, are allowable within the
  standard model.

  The $B_{c}^{\ast}$ meson has a relatively large mass. In addition, both
  constituent quarks $b$ and $c$ of the $B_{c}^{\ast}$ meson can decay individually.
  Therefore, the $B_{c}^{\ast}$ meson has rich weak decay channels.
  The $B_{c}^{\ast}$ meson weak decays, similar to the pseudoscalar
  $B_{c}$ meson weak decays \cite{zpc51,prd49,usp38,prd77.074013,prd89.114019,ahep2015.104378,qwg},
  can be divided into three classes:
  (1) the $c$ quark decay with the spectator $b$ quark,
  (2) the $b$ quark decay with the $c$ quark as a spectator,
  and (3) the $b$ and $c$ quarks annihilation into a virtual
  $W^{\pm}$ boson.
  This property makes the $B_{c}^{\ast}$ meson another
  potentially fruitful laboratory for studying the weak decay
  mechanism of heavy flavor hadrons.

  The study of $B_{c}^{\ast}$ weak decays might be interesting,
  but has not really started yet.
  One of the major reasons is the extraordinary difficulty of
  producing the $B_{c}^{\ast}$ meson.
  The production cross section for the $B_{c}^{\ast}$ meson
  in hadronic collisions via the dominant process of $g$ $+$ $g$
  ${\to}$ $B_{c}^{\ast}$ $+$ $b$ $+$ $\bar{c}$
  \cite{qwg,plb355,plb364,prd54.4344,epjc38.267,prd72.114009}
  is at least at the order of ${\alpha}_{s}^{4}$.
  The nature of QCD's asymptotic freedom implies a much small possibility
  of creating two heavy quark pairs ($b\bar{b}$ and $c\bar{c}$) from
  the vacuum at the ultrahigh energy.
  Fortunately, the high luminosities of the running LHC and the
  future {\em Super proton proton Collider} (S$pp$C,
  which is still under discussion today) will promisingly improve
  this situation. It is expected that a huge amount of the
  $B_{c}^{\ast}$ data samples would be accumulated, and offer
  a valuable opportunity to investigate the $B_{c}^{\ast}$ weak
  decays.

  As is well known, there exist some hierarchical structures among the
  Cabibbo-Kobayashi-Maskawa (CKM) matrix elements.
  The CKM coupling strength for the bottom quark weak decay is proportional to
  ${\vert}V_{cb}{\vert}$ ${\sim}$ ${\cal O}({\lambda}^{2})$ or
  ${\vert}V_{ub}{\vert}$ ${\sim}$ ${\cal O}({\lambda}^{3})$,
  while the CKM coupling strength for the charm quark weak decay is proportional to
  ${\vert}V_{cs}{\vert}$ ${\sim}$ ${\cal O}(1)$ or
  ${\vert}V_{cd}{\vert}$ ${\sim}$ ${\cal O}({\lambda})$,
  with the Wolfenstein parameter ${\lambda}$ ${\approx}$ $0.2$ \cite{pdg}.
  The $B_{q}$ ($q$ $=$ $u$, $d$, $s$) weak decays
  are induced dominantly by the bottom quark decay
  with the phenomenological spectator scheme.
  The $B_{c}^{\ast}$ ${\to}$ $B_{q}V$, $B_{q}P$ decays
  are actually induced  by
  the charm quark weak decay, where $V$ and $P$ denote respectively
  the lightest 9-pelts $SU(3)$ vector and pseudoscalar mesons.
  With respect to the $B_{q}$ weak decays, the $B_{c}^{\ast}$
  ${\to}$ $B_{q}V$, $B_{q}P$ decays are favored by the CKM matrix elements.
  In this paper, we will study the $B_{c}^{\ast}$ ${\to}$ $B_{u,d,s}V$, $B_{u,d,s}P$ weak
  decays with the QCD factorization (QCDF) approach
  \cite{prl83.1914,npb591.313,npb606.245,plb488.46,plb509.263,
  prd64.014036,npb774.64,npb832.109,plb750.348},
  in order to provide an available reference for the
  future experimental investigation.
  There is a more than $2.0\,{\sigma}$ discrepancy between the value for
  CKM matrix element ${\vert}V_{cs}{\vert}$ obtained from semileptonic $D$ decays
  and that from leptonic $D_{s}$ decays\footnotemark[1] \cite{pdg}.
  The $B_{c}^{(\ast)}$ ${\to}$ $B_{s}V$, $B_{s}P$ decays, together with
  semileptonic $D$ decays and leptonic $D_{s}$ decays, will provide
  ${\vert}V_{cs}{\vert}$ with more stringent constraints.
  \footnotetext[1]{The value for CKM matrix element ${\vert}V_{cs}{\vert}$
  is ${\vert}V_{cs}{\vert}$ $=$ $0.953{\pm}0.008{\pm}0.024$ from semileptonic
  $D$ decays, and ${\vert}V_{cs}{\vert}$ $=$ $1.008{\pm}0.021$ from
  leptonic $D_{s}$ decays \cite{pdg}.}
  In addition, some of the $B_{c}$ weak decays,
  for example, the $B_{c}$ ${\to}$ $B_{s}{\pi}$ decay \cite{prl.111},
  have been measured now. One possible background might come from
  the $B_{c}^{\ast}$ decays, due to a slightly larger production cross
  section ${\sigma}(B_{c}^{\ast})$ than ${\sigma}(B_{c})$ in hadronic
  collisions \cite{plb364,prd54.4344,epjc38.267,prd72.114009},
  and the nearly equal mass $m_{B_{c}^{\ast}}$ ${\simeq}$
  $m_{B_{c}}$ \cite{prd86.094510}.
  Hence, the study of the $B_{c}^{\ast}$ ${\to}$ $B_{u,d,s}V$, $B_{u,d,s}P$
  decays will be helpful to the experimental analysis
  on the $B_{c}$ ${\to}$ $B_{u,d,s}V$, $B_{u,d,s}P$ decays.

  This paper is organized as follows.
  The theoretical framework and decay amplitudes will be presented
  in Section \ref{sec02}.
  Section \ref{sec03} is the numerical results and discussion.
  The last section is a summary.

  \section{theoretical framework}
  \label{sec02}
  \subsection{The effective Hamiltonian}
  \label{sec0201}
  Using the operator product expansion and the renormalization
  group (RG) method, the low-energy effective weak Hamiltonian describing
  the $B_{c}^{\ast}$ ${\to}$ $B_{u,d,s}V$, $B_{u,d,s}P$ decays
  has the following general structure \cite{9512380},
   \begin{equation}
  {\cal H}_{\rm eff}\, =\, \frac{G_{F}}{\sqrt{2}}
   \sum\limits_{q,q^{\prime}=s,d} V_{cq}^{\ast} V_{uq^{\prime}}
   \Big\{  C_{1}({\mu})\, Q_{1}({\mu})\,
         + C_{2}({\mu})\, Q_{1}({\mu}) \Big\}
         + {\rm h.c.}
   \label{hamilton},
   \end{equation}
  where the Fermi coupling constant $G_{F}$ ${\simeq}$
  $1.166{\times}10^{-5}\,{\rm GeV}^{-2}$ \cite{pdg};
  $V_{cq}^{\ast} V_{uq^{\prime}}$ is a product of the CKM
  matrix elements. Using the Wolfenstein parameterization,
  there are \cite{pdg}
   \begin{equation}
   V_{cs}^{\ast}V_{ud}\ =\ 1 - {\lambda}^{2} - \frac{1}{2}\,A^{2}\,{\lambda}^{4}
   + \frac{1}{2}\,A^{2}\,{\lambda}^{6}\,\big\{ 1 -{\rho}^{2}-{\eta}^{2} -2\,
   ({\rho}-i\,{\eta}) \big\} + {\cal O}({\lambda}^{8})
   \label{vcs-vud},
   \end{equation}
   \begin{equation}
   V_{cs}^{\ast}V_{us}\ =\ {\lambda} - \frac{{\lambda}^{3}}{2}
    - \frac{{\lambda}^{5}}{8} - \frac{{\lambda}^{7}}{16}
    - \frac{1}{2}\,A^{2}\,{\lambda}^{5}
    + \frac{1}{2}\,A^{2}\,{\lambda}^{7}\, \big\{
      \frac{1}{2} - {\rho}^{2}-{\eta}^{2}
    -2\,({\rho}-i\,{\eta}) \big\} + {\cal O}({\lambda}^{8})
   \label{vcs-vus},
   \end{equation}
   \begin{equation}
   V_{cd}^{\ast}V_{ud}\ =\ -V_{cs}^{\ast}V_{us}
   -A^{2}\,{\lambda}^{5}\,({\rho}-i\,{\eta}) + {\cal O}({\lambda}^{8})
   \label{vcd-vud},
   \end{equation}
   \begin{equation}
   V_{cd}^{\ast}V_{us}\ =\ - {\lambda}^{2}
   + \frac{1}{2}\,A^{2}\,{\lambda}^{6}\,\big\{
     1-2\,({\rho}-i\,{\eta}) \big\} + {\cal O}({\lambda}^{8})
   \label{vcd-vus},
   \end{equation}
  where the values for these Wolfenstein parameters $A$, ${\lambda}$,
  ${\rho}$ and ${\eta}$ are given in Table \ref{tab:input}.

  The renormalization scale ${\mu}$
  separates the physical contributions into two parts.
  The hard contributions above the scale ${\mu}$ are summarized
  into the Wilson coefficients $C_{i}(\mu)$.
  With the RG equation for $C_{i}(\mu)$, the Wilson coefficients
  at an appropriate scale ${\mu}_{c}$ ${\sim}$ ${\cal O}(m_{c})$ for
  the charm quark decay are given by \cite{9512380}
  \begin{equation}
  \vec{C}({\mu}_{c})\, =\, U_{4}({\mu}_{c},m_{b})\,U_{5}(m_{b},m_{W})\,\vec{C}(m_{W})
  \label{ci},
  \end{equation}
  where $m_{W}$, $m_{b}$ and $m_{c}$ are the mass of the $W$ boson,
  $b$ quark and $c$ quark, respectively.
  Here $U_{f}(m_{2},m_{1})$ denotes the RG evolution matrix
  for $f$ active flavors.
  The initial values for the Wilson coefficients $\vec{C}(m_{W})$
  at scale ${\mu}_{W}$ $=$ $m_{W}$ to a desired order in
  ${\alpha}_{s}$ can be calculated with perturbation
  theory. The expressions for the RG evolution matrix $U_{f}(m_{2},m_{1})$
  and Wilson coefficients $\vec{C}(m_{W})$, including both leading order
  (LO) and next-to-leading order (NLO) corrections,
  have been presented in Ref.\cite{9512380}.
  The contributions below the scale ${\mu}$ are included in
  the hadronic matrix elements (HME) where the local four-quark
  operators $Q_{i}$ are sandwiched between the initial and final states.
  The expressions for the four-quark operators in question are
   \begin{equation}
   Q_{1}\, =\,
   \big[ \bar{q}_{\alpha}\,{\gamma}_{\mu}\,(1-{\gamma}_{5})\,c_{\alpha} \big]\,
   \big[ \bar{u}_{\beta}\, {\gamma}^{\mu}\,(1-{\gamma}_{5})\,q^{\prime}_{\beta} \big]
   \label{q1},
   \end{equation}
   \begin{equation}
   Q_{2}\, =\,
   \big[ \bar{q}_{\alpha}\,{\gamma}_{\mu}\,(1-{\gamma}_{5})\,c_{\beta} \big]\,
   \big[ \bar{u}_{\beta}\, {\gamma}^{\mu}\,(1-{\gamma}_{5})\,q^{\prime}_{\alpha} \big]
   \label{q2},
   \end{equation}
  where the subscripts ${\alpha}$ and ${\beta}$ are color indices.
  It should be pointed out that
  (1)
  because the contributions from the penguin operators and annihilation
  topologies are proportional to the CKM factor $V_{cb}^{\ast}V_{ub}$
  ${\sim}$ ${\cal O}({\lambda}^{5})$ and therefore negligible in the
  actual calculation of branching ratio \cite{prd89.114019}, only the
  contributions of tree operators are considered here.
  (2)
  The participation of the strong interaction, especially, the
  nonperturbative QCD effects, makes the theoretical treatment of HME
  very complicated.
  The main problem at this stage is how to effectively factorize HME
  into hard and soft parts, and how to evaluate HME properly.

  \subsection{Hadronic matrix elements}
  \label{sec0202}
  Hadronic matrix elements might be the most intricate part
  in the calculation of heavy flavor weak decay, due to the
  entanglement of perturbative and nonperturbative contributions.
  Phenomenologically, one has to turn to some approximation
  and assumption, which bring uncertainties and model dependence
  to theoretical predictions.
  A simple approximation is the naive factorization ansatz (NF)
  according to Bjorken's color transparency argument, which says that the colorless
  energetic hadron has flown away from the weak interaction point during the
  formation time of the emission hadron \cite{npb11.325}. With the NF approach,
  HME is parameterized as a product of decay constants
  and hadron transition form factors \cite{plb73.418,npb133.315,zpc29.637,zpc34.103}.
  A major flaw of the NF approach is the disappearance of scale dependence
  and strong phases from HME, which results directly in a scale-sensitive
  nonphysical prediction and none of $CP$ violation for nonleptonic meson weak
  decays.
  In order to overcome these shortcomings of the NF approach,
  nonfactorizable contributions to HME should be carefully considered,
  as commonly recognized.
  Some QCD-inspired models, such as, the QCDF approach \cite{prl83.1914,
  npb591.313,npb606.245,plb488.46,plb509.263,prd64.014036,npb774.64,npb832.109,plb750.348},
  the soft and collinear effective theory \cite{prd63.014006,
  prd63.114020,plb516.134,prd65.054022,prd66.014017,npb643.431,
  plb553.267,npb685.249}, the perturbative QCD approach \cite{pqcd1,pqcd2,pqcd3},
  and so on, have been developed recently, based on the Lepage-Brodsky treatment
  on exclusive processes \cite{prd22} and some power counting rules in the
  expansion in ${\alpha}_{s}$ and ${\Lambda}_{\rm QCD}/m_{Q}$,
  where ${\alpha}_{s}$ is the strong coupling, ${\Lambda}_{\rm QCD}$ is the
  QCD characteristic scale, and $m_{Q}$ is the mass of a heavy quark.
  In these QCD-inspired models, HME is generally written
  as a convolution integral of hadron's distribution amplitudes (DAs)
  and hard rescattering kernels.
  A virtue of the QCDF approach is that the NF's result can be reproduced, if both
  the nonfactorizable contributions and the power suppressed contributions are
  neglected \cite{prl83.1914,npb591.313,npb606.245,plb488.46,plb509.263,prd64.014036}.

  For the $B_{c}^{\ast}$ ${\to}$ $B_{q}V$, $B_{q}P$ decays
  ($q$ $=$ $u$, $d$, $s$), the spectator quark is a heavy quark
  --- the bottom quark.
  It is generally assumed that the bottom quark in both the $B_{c}^{\ast}$
  and $B_{q}$ mesons is nearly on shell, and that the gluon exchanged
  between the heavy spectator quark and other quarks is soft.
  The virtuality of emission gluon from the spectator quark is
  of order ${\Lambda}_{\rm QCD}^{2}$.
  The contributions of spectator scattering are power suppressed relative to
  the leading order contributions \cite{npb591.313}.
  In addition,
  it is supposed that the recoiled $B_{q}$ meson should move slowly
  in the rest frame of the $B_{c}^{\ast}$ meson. There should be a
  large overlap between the $B_{c}^{\ast}$ and $B_{q}$ mesons.
  The recoiled $B_{q}$ meson cannot be clearly factorized from the
  $B_{c}^{\ast}B_{q}$ system due to the soft and nonperturbative
  contributions.  The $B_{c}^{\ast}B_{q}$ system should
  be parameterized by some physical from factors.
  Hence, with the QCDF approach, up to leading power corrections of order
  ${\Lambda}_{\rm QCD}/m_{Q}$, hadronic matrix elements have the
  following structure \cite{npb591.313},
  \begin{equation}
 {\langle}B_{q}M{\vert}Q_{i}{\vert}B_{c}^{\ast}{\rangle}\, =\,
  f_{M} \sum_{j} F^{B_{c}^{\ast}{\to}B_{q}}_{j}
 {\int}dx\, {\cal H}_{ij}(x)\, {\phi}(x)\, =\,
  f_{M} \sum_{j} F^{B_{c}^{\ast}{\to}B_{q}}_{j}\,
  \Big\{ 1+{\alpha}_{s}\,r_{j}+{\cdots} \Big\}
  \label{hadronic},
  \end{equation}
  where $f_{M}$ is the decay constant for the light final $M$ (${\equiv}$
  $V$ and $P$) meson; $F^{B_{c}^{\ast}{\to}B_{q}}_{j}$ is a transition
  form factor; ${\cal H}_{ij}(x)$ is a hard rescattering kernel;
  ${\phi}(x)$ is a DA of parton momentum fraction $x$. 
  For the light pseudoscalar $P$ and longitudinally polarized vector $V$ mesons,
  the leading twist DAs are expanded in terms of the Gegenbauer
  polynomials \cite{jhep.0605.004,jhep.0703.069}
  \begin{equation}
 {\phi}_{P}(x)\ =\ 6\,x\,\bar{x}\,
  \Big\{ 1+ \sum\limits_{n=1}
   a_{n}^{P}\, C_{n}^{3/2}(x-\bar{x}) \Big\}
  \label{das-p-twist2-a},
  \end{equation}
  \begin{equation}
 {\phi}_{V}(x) \, =\, 6\,x\,\bar{x}\,
  \Big\{ 1+ \sum\limits_{n=1}
   a_{n}^{V}\, C_{n}^{3/2}(x-\bar{x}) \Big\}
  \label{da-v-twist2-v},
  \end{equation}
  where $\bar{x}$ $=$ $1$ $-$ $x$; $a_{n}^{P,V}$ is a nonperturbative
  parameter, also called the Gegenbauer moment. The expressions for
  the Gegenbauer polynomials $C_{n}^{3/2}(z)$ are
  \begin{equation}
  C_{1}^{3/2}(z) = 3\,z,
  \quad
  C_{2}^{3/2}(z) = \frac{3}{2}\,(5\,z^{2}-1),
  \quad
  {\cdots}
  \label{Gegenbauer-polynomials}
  \end{equation}

  \begin{figure}[ht]
  \includegraphics[width=0.98\textwidth,bb=85 645 525 720]{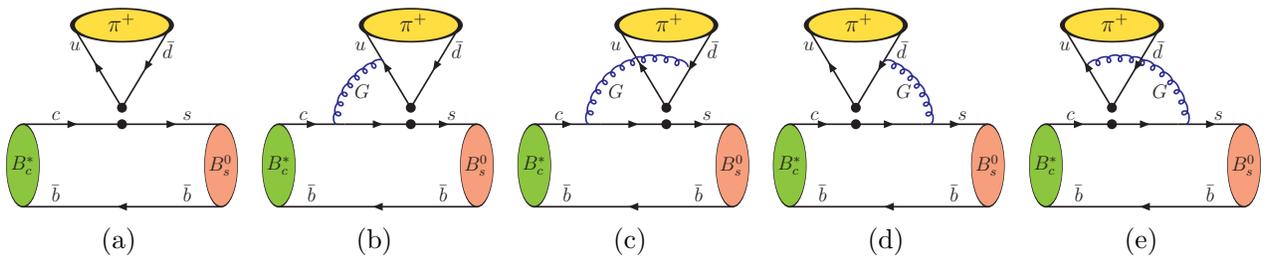}
  \caption{Feynman diagrams for $B_{c}^{{\ast}}$ ${\to}$ $B_{s}{\pi}$
  decay within the QCDF framework, where (a) denotes the
  factorizable contributions, and (b,c,d,e) correspond to the nonfactorizable
  vertex corrections at the order of ${\alpha}_{s}$.}
  \label{feynman}
  \end{figure}

  \subsection{Decay amplitudes}
  \label{sec0203}
  The typical Feynman diagrams for the $B_{c}^{{\ast}}$ ${\to}$ $B_{s}{\pi}$ decay
  within the QCDF framework are shown in Fig.\ref{feynman}, where no hard
  gluons are exchanged between the spectator quark and other partons.
  There is no gluon exchange in factorizable topology of Fig.\ref{feynman}(a),
  so the emitted hadron matrix element is entirely separated from that of the
  $B_{c}^{{\ast}}B_{s}$ system. In this approximation,
  the hard rescattering kernel ${\cal H}_{ij}$ $=$ $1$ and the integral in
  Eq.(\ref{hadronic}) reduces to the normalization condition for distribution amplitude.
  According to the QCDF power counting rules, the leading order contributions
  come from the factorizable topology of Fig.\ref{feynman}(a), and
  recover the NF's results at the order of ${\alpha}_{s}^{0}$.
  For the radiative correction diagrams in Fig.\ref{feynman}(b-e),
  hard gluons are exchanged between the emission meson and the
  $B_{c}^{{\ast}}B_{s}$ system.
  The hard rescattering kernel ${\cal H}_{ij}$ and $x$-integral in Eq.(\ref{hadronic})
  are nontrivial.
  It has already been shown \cite{npb591.313,npb606.245,plb488.46,
  plb509.263,prd64.014036} that although both collinear and soft
  divergences exist for each of diagrams in Fig.\ref{feynman}(b-e), infrared
  divergences cancel after summing up the vertex corrections.
  The strong phases could then come from HME. The renormalization
  scale ${\mu}$ dependence of HME is recuperated from
  the nonfactorizable contributions, which will reduce partly
  the ${\mu}$-dependence of Wilson coefficients.

  After a straightforward calculation using the QCDF master formula
  Eq.(\ref{hadronic}), the amplitudes
  for the $B_{c}^{{\ast}}$ ${\to}$ $B_{q}M$ decays ($q$ $=$ $u$, $d$, $s$) are written as
   \begin{equation}
  {\cal A}(B_{c}^{\ast}{\to}B_{q}M)\, =\,
  {\langle}B_{q}M{\vert}{\cal H}_{\rm eff}{\vert}B_{c}^{\ast}{\rangle}\, =\,
   \frac{G_{F}}{\sqrt{2}}\, V_{cq}^{\ast} V_{uq^{\prime}}\,
   a_{i}\, {\langle}M{\vert}j^{\mu}{\vert}0{\rangle}\,
   {\langle}B_{q}{\vert} j_{\mu} {\vert}B_{c}^{\ast}{\rangle}
   \label{amp-bc-b-m}.
   \end{equation}
  With the naive dimensional regularization scheme,
  the effective coefficients  are \cite{npb591.313,npb606.245,plb488.46,plb509.263,prd64.014036}:
  \begin{eqnarray}
   a_{1}
   &=& C_{1}^{\rm NLO}+\frac{1}{N_{c}}\,C_{2}^{\rm NLO}
    + \frac{{\alpha}_{s}}{4{\pi}}\, \frac{C_{F}}{N_{c}}\,
      C_{2}^{\rm LO}\, {\cal V}
   \label{a1}, \\
   a_{2}
   &=& C_{2}^{\rm NLO}+\frac{1}{N_{c}}\,C_{1}^{\rm NLO}
    + \frac{{\alpha}_{s}}{4{\pi}}\, \frac{C_{F}}{N_{c}}\,
      C_{1}^{\rm LO}\, {\cal V}
   \label{a2},
  \end{eqnarray}
  \begin{equation}
  {\cal V} = 6\,{\log}\Big( \frac{m_{c}^{2}}{{\mu}^{2}} \Big)
    -  18 - \Big( \frac{1}{2}+i\,3\,{\pi} \Big)
    +  \Big( \frac{11}{2}-i\,3\,{\pi} \Big)\,a_{1}^{M}
    -   \frac{21}{20}\,a_{2}^{M} +{\cdots}
  \label{vc},
  \end{equation}
  where $N_{c}$ $=$ $3$ and $C_{F}$ $=$ $4/3$;
  $C_{1,2}^{\rm NLO,LO}$ are Wilson coefficients
  containing NLO or LO contributions; $a_{i}^{M}$ is a
  Gegenbauer moment. For the transversely polarized
  vector meson, the vertex factor ${\cal V}$ $=$ $0$ beyond
  the leading twist DAs.
  For convenience, the numerical values for $a_{1,2}$
  of the $B_{c}^{\ast}$ ${\to}$ $B_{q}{\pi}$ decay
  are listed in Table \ref{tab:ai}.

  There are some comments on the coefficients $a_{1,2}$.
  (1)
  The first two terms on the right hand side of Eq.(\ref{a1})
  and Eq.(\ref{a2}) correspond to the leading order contributions.
  The third terms correspond to nonfactorizable contributions.
  The NF scenario follows when one neglects the
  nonfactorizable contributions, i.e., ${\cal V}$ $=$ $0$.
  (2)
  Nonfactorizable vertex corrections to HME are of order ${\alpha}_{s}$.
  They include the dependence on the renormalization scale.
  It is shown \cite{prd64.014036} that with the RG equations
  for the Wilson coefficients at leading order logarithm
  approximation, one can obtain
  \begin{math} \displaystyle
  {\mu}\frac{{\rm d}}{{\rm d}{\mu}}a_{1,2}\, =\, 0
  \end{math}.
  In principle, the residual scale dependence could be compensated by higher
  order corrections to HME.
  (3)
  Compared with the LO contributions,
  nonfactorizable contributions are generally suppressed  by
  ${\alpha}_{s}$ and the factor $1/N_{c}$ (see Eq.(\ref{a1}) and Eq.(\ref{a2})).
  Because the LO contributions of $a_{2}$
  are color-suppressed, vertex corrections multiplied by
  the large Wilson coefficient $C_{1}^{\rm LO}$ could be
  sizable to branching rates of the $a_{2}$-dominated heavy flavor decays.
  The coefficients $a_{1,2}$ contain strong phases via the
  imaginary parts of vertex corrections.
  Correspondingly, strong scattering phase of
  $a_{1}$ ($a_{2}$) is small (large).
  This argument is also confirmed by the numerical results for
  $a_{1,2}$ in Table \ref{tab:ai}.
  (4)
  With the QCDF approach, nonfactorizable radiative corrections
  to HME occur first at order ${\alpha}_{s}$
  as well as the leading strong phases at order ${\alpha}_{s}$.
  In addition, it should be pointed out that nonfactorizable
  power corrections beyond leading order are neglected here.
  For the charm quark decay, power ${\Lambda}_{\rm QCD}/m_{c}$ is
  comparable to ${\alpha}_{s}$.
  The strong phases due to soft (hard) interactions
  are of order ${\Lambda}_{\rm QCD}/m_{c}$ (${\alpha}_{s}$).
  One should not expect these phases to have great
  precision, as stated in Ref.\cite{npb591.313}.
  (5)
  With the QCDF approach, the values for $a_{1,2}$ are
  close to those for the charm quark decay
  \cite{npb268.16,plb252.690,ijmpa14.937,epjc55.607,prd81.074021},
  ${\vert}a_{1,2}{\vert}$ ${\approx}$ ${\vert}C_{1,2}{\vert}$, and
  basically consistent with those of the large-$N_{c}$ approach \cite{npb268.16}.

   \begin{table}[ht]
   \caption{The numerical values for the Wilson coefficients and
   $a_{1,2}$ for the $B_{c}^{\ast}$ ${\to}$ $B_{q}{\pi}$ decay.}
   \label{tab:ai}
   \begin{ruledtabular}
   \begin{tabular}{c|cc|cc|cc|cc}
   scale
  & \multicolumn{2}{c|}{LO}
  & \multicolumn{2}{c|}{NLO}
  & \multicolumn{2}{c|}{NF}
  & \multicolumn{2}{c}{QCDF}
    \\ \cline{2-3} \cline{4-5} \cline{6-7} \cline{8-9}
    ${\mu}$
  & $C_{1}$ & $C_{2}$ & $C_{1}$ & $C_{2}$ & $a_{1}$ & $a_{2}$ & $a_{1}$ & $a_{2}$ \\ \hline
    $0.8\,m_{c}$
  & $1.310$ & $-0.553$ & $1.253$ & $-0.473$ & $1.096$ & $-0.055$
  & $1.235+i\,0.081$ & $-0.384-i\,0.192$  \\
    $m_{c}$
  & $1.259$ & $-0.479$ & $1.209$ & $-0.404$ & $1.074$ & $-0.001$
  & $1.193+i\,0.060$ & $-0.313-i\,0.157$ \\
    $1.2\,m_{c}$
  & $1.227$ & $-0.430$ & $1.180$ & $-0.358$ & $1.061$ & $~0.035$
  & $1.167+i\,0.048$ & $-0.269-i\,0.137$
   \end{tabular}
   \end{ruledtabular}
   \end{table}

  The hadronic matrix elements of diquark current operators are defined as
  \cite{zpc29.637}:
   \begin{equation}
  {\langle}V({\epsilon},p){\vert}\bar{q}_{1}\,{\gamma}^{\mu}\,(1-{\gamma}_{5})\,q_{2}
  {\vert}0{\rangle}\, =\, f_{V}\,m_{V}\,{\epsilon}^{{\ast}{\mu}}
   \label{vector},
   \end{equation}
   \begin{equation}
  {\langle}P(p){\vert}\bar{q}_{1}\,{\gamma}^{\mu}\,(1-{\gamma}_{5})\,q_{2}
  {\vert}0{\rangle}\, =\, -i\,f_{P}\,p^{\mu}
   \label{pseudoscalar},
   \end{equation}
    \begin{eqnarray}
   & &
   {\langle}B_{q}(p_{2}){\vert}\bar{q}\,{\gamma}_{\mu}\,(1-{\gamma}_{5})\,c
   {\vert}B_{c}^{\ast}(p_{1},{\epsilon}){\rangle}
    \nonumber \\ &=&
  -{\varepsilon}_{{\mu}{\nu}{\alpha}{\beta}}\,
   {\epsilon}^{{\nu}}\, q^{\alpha}\, (p_{1}+p_{2})^{\beta}\,
     \frac{V(q^{2})}{m_{B_{c}^{\ast}}+m_{B_{q}}}
   -i\,2\,m_{B_{c}^{\ast}}\,\frac{{\epsilon}{\cdot}q}{q^{2}}\,
    q_{\mu}\, A_{0}(q^{2})
    \nonumber \\ & &
    -i\,{\epsilon}_{\mu}\, ( m_{B_{c}^{\ast}}+m_{B_{q}} )\, A_{1}(q^{2})
   -i\,\frac{{\epsilon}{\cdot}q}{m_{B_{c}^{\ast}}+m_{B_{q}}}\,
   ( p_{1} + p_{2} )_{\mu}\, A_{2}(q^{2})
    \nonumber \\ & &
   +i\,2\,m_{B_{c}^{\ast}}\,\frac{{\epsilon}{\cdot}q}{q^{2}}\,
   q_{\mu}\, A_{3}(q^{2})
    \label{a0123v},
    \end{eqnarray}
  where $f_{V}$ and $f_{P}$ are the decay constants of
  vector $V$ and pseudoscalar $P$ mesons, respectively;
  $q$ $=$ $p_{1}$ $-$ $p_{2}$;
  ${\epsilon}$ is the polarization vector of vector mesons;
  $V(q^{2})$ and $A_{0,1,2,3}(q^{2})$ are the $B_{c}^{\ast}$ ${\to}$ $B_{q}$ transition
  form factors.
  To eliminate singularities at the pole of $q^{2}$ $=$ $0$,
  a relation, $A_{0}(0)$ $=$ $A_{3}(0)$, is
  required, with $A_{3}(q^{2})$ given by \cite{zpc29.637}:
  \begin{equation}
  2\,m_{B_{c}^{\ast}}\,A_{3}(q^{2})\,=\,
  (m_{B_{c}^{\ast}}+m_{B_{q}})\,A_{1}(q^{2})
 +(m_{B_{c}^{\ast}}-m_{B_{q}})\,A_{2}(q^{2})
  \label{form01}.
  \end{equation}

  In the bottom conservation transition $B_{c}^{\ast}$ ${\to}$ $B_{q}$,
  both the initial and final mesons contain a heavy bottom quark.
  After a sudden kick, the $B_{q}$ meson would move slowly,
  even remain nearly intact, with respect to the $B_{c}^{\ast}$ meson.
  Therefore, the zero-recoil configuration ($q^{2}$ $=$ $0$) would be
  a good approximation.
  Simultaneously, the emission meson would take up most of the
  energy available and fly rapidly away from the interaction point.
  This fact not only reproduces the NF scenario (Fig.\ref{feynman}(a))
  but also requires the exchanged gluon in vertex corrections
  (Fig.\ref{feynman}(b-e)) to be hard. Due to the large virtuality of
  gluon exchanged between the emitted light meson and the
  $B_{c}^{\ast}B_{q}$ system,
  perturbative calculation of nonfactorizable vertex corrections with
  the QCDF approach should be applicable and reliable.

  With the form factors given above, the decay amplitudes
  are expressed as
  \begin{eqnarray}
  {\cal A}(B_{c}^{\ast}{\to}B_{q}V) &=&
  -i\, \frac{G_{F}}{\sqrt{2}}\, m_{V}\, f_{V}\,V_{cq}^{\ast} V_{uq^{\prime}}\,
   \big\{ a_{1}\,{\delta}_{B_{q},B_{d,s}}
  + a_{2}\,{\delta}_{B_{q},B_{u}} \big\}\,
   \big\{ ({\epsilon}_{B_{c}^{\ast}}{\cdot}{\epsilon}_{V}^{\ast})\,
   ( m_{B_{c}^{\ast}}+m_{B_{q}} )\, A_{1}
   \nonumber \\ & &
  +\, ({\epsilon}_{B_{c}^{\ast}}{\cdot}p_{V})\,
   (p_{B_{c}^{\ast}}{\cdot}{\epsilon}_{V}^{\ast})\,
   \frac{2\,A_{2}}{m_{B_{c}^{\ast}}+m_{B_{q}}}
  +i\, {\epsilon}_{{\mu}{\nu}{\alpha}{\beta}}\,
   {\epsilon}_{B_{c}^{\ast}}^{\mu}\,
   {\epsilon}_{V}^{{\ast}{\nu}}\,p_{B_{c}^{\ast}}^{\alpha}\,
   p_{V}^{\beta}\, \frac{2\,V}{m_{B_{c}^{\ast}}+m_{B_{q}}} \big\}
  \label{amp-bv-00},
  \end{eqnarray}
  \begin{equation}
  {\cal A}(B_{c}^{\ast}{\to}B_{q}P)\, =\,
   \sqrt{2}\, G_{F}\, m_{B_{c}^{\ast}}\,
  ({\epsilon}_{B_{c}^{\ast}}{\cdot}p_{B_{q}})\, f_{P}\,
   A_{0}\, V_{cq}^{\ast} V_{uq^{\prime}}\,
   \big\{ a_{1}\,{\delta}_{B_{q},B_{d,s}}
  + a_{2}\,{\delta}_{B_{q},B_{u}} \big\}
  \label{amp-bp-00}.
  \end{equation}

  The $B_{c}^{\ast}$ ${\to}$ $B_{q}V$ decay amplitude
  is a sum of $S$-, $P$-, $D$-wave amplitudes \cite{prd39.3339,prd45.193}, i.e.,
  \begin{equation}
 {\cal A}(B_{c}^{\ast}{\to}B_{q}V)=
  a\,({\epsilon}_{B_{c}^{\ast}}{\cdot}{\epsilon}_{V}^{\ast})
  + \frac{b}{m_{B_{c}^{\ast}}\,m_{V} }
   ({\epsilon}_{B_{c}^{\ast}}{\cdot}p_{V})\,
   (p_{B_{c}^{\ast}}{\cdot}{\epsilon}_{V}^{\ast})
  + \frac{i\,c}{m_{B_{c}^{\ast}}\,m_{V} }
   {\epsilon}_{{\mu}{\nu}{\alpha}{\beta}}\,
   {\epsilon}_{B_{c}^{\ast}}^{\mu}\,
   {\epsilon}_{V}^{{\ast}{\nu}}\,p_{B_{c}^{\ast}}^{\alpha}\, p_{V}^{\beta}
   \label{amp-bv-01},
  \end{equation}
  with $a$, $b$, $c$, the $S$-, $D$- and $P$-wave amplitudes respectively,
  in the notation of \cite{prd45.193},
  \begin{equation}
  a\, =\, {\cal F}\, ( m_{B_{c}^{\ast}}+m_{B_{q}} )\, A_{1}
  \label{amp-bv-a},
  \end{equation}
  \begin{equation}
  b\, =\, {\cal F}\, \frac{2\,m_{B_{c}^{\ast}}\,m_{V}}{m_{B_{c}^{\ast}}+m_{B_{q}}}\,A_{2}
  \label{amp-bv-b},
  \end{equation}
  \begin{equation}
  c\, =\, {\cal F}\, \frac{2\,m_{B_{c}^{\ast}}\,m_{V}}{m_{B_{c}^{\ast}}+m_{B_{q}}}\,V
  \label{amp-bv-c},
  \end{equation}
  \begin{equation}
  {\cal F}\, =\, -i\, \frac{G_{F}}{\sqrt{2}}\, m_{V}\, f_{V}\,V_{cq}^{\ast} V_{uq^{\prime}}\,
   \big\{ a_{1}\,{\delta}_{B_{q},B_{d,s}} + a_{2}\,{\delta}_{B_{q},B_{u}} \big\}
  \label{amp-bv-f}.
  \end{equation}
  From the above expressions, one can find that the $P$- and $D$-wave
  amplitudes are suppressed by a factor of
  $\frac{2\,m_{B_{c}^{\ast}}\,m_{V}}{(m_{B_{c}^{\ast}}+m_{B_{q}})^{2}}$
  relative to the $S$-wave amplitude.
  The relations among the helicity amplitudes and the $S$-, $P$-,
  $D$-wave amplitudes are \cite{prd45.193}
   \begin{equation}
  H_{\pm}\ =\ a\,{\pm}\,c\,\sqrt{y^{2}-1}
   \label{amp-bv-hpm},
   \end{equation}
   \begin{equation}
  H_{0}\ =\ -a\,y-b\,(y^{2}-1)
   \label{amp-bv-h0},
   \end{equation}
   \begin{equation}
  y\ =\ \frac{ p_{B_{c}^{\ast}}{\cdot}p_{V} }{ m_{B_{c}^{\ast}}\,m_{V} }
   \ =\ \frac{ m_{B_{c}^{\ast}}^{2}-m_{B_{q}}^{2}+m_{V}^{2} }{ 2\,m_{B_{c}^{\ast}}\,m_{V} }
   \label{amp-bv-x},
   \end{equation}
   \begin{equation}
  p_{\rm cm}\ =\ \frac{ \sqrt{ [m_{B_{c}^{\ast}}^{2}-(m_{B_{q}}+m_{V})^{2}]\,
   [m_{B_{c}^{\ast}}^{2}-(m_{B_{q}}-m_{V})^{2}] } }{ 2\,m_{B_{c}^{\ast}} }
   \label{amp-bv-p},
   \end{equation}
   \begin{equation}
  p_{\rm cm}^{2}\ =\ m_{V}^{2}\,(y^{2}-1)
   \label{amp-bv-px},
   \end{equation}
  where $p_{\rm cm}$ is the common momentum of final states
  in the rest frame of the $B_{c}^{\ast}$ meson.

  We assume that the vector mesons are ideally mixed in the
  singlet-octet basis, i.e., ${\phi}$ $=$ $s\bar{s}$
  and ${\omega}$ $=$ $(u\bar{u}+d\bar{d})/\sqrt{2}$.
  As for the pseudoscalar ${\eta}$ and
  ${\eta}^{\prime}$ mesons, they are usually written as a linear
  superposition of states in either flavor basis or the
  singlet-octet basis.
  Here, we adopt the quark flavor basis description proposed in Ref.
  \cite{prd.58.114006}, i.e.,
   \begin{equation}
   \left(\begin{array}{c}
  {\eta} \\ {\eta}^{\prime}
   \end{array}\right) =
   \left(\begin{array}{cc}
  {\cos}{\phi} & -{\sin}{\phi} \\
  {\sin}{\phi} &  {\cos}{\phi}
   \end{array}\right)
   \left(\begin{array}{c}
  {\eta}_{q} \\ {\eta}_{s}
   \end{array}\right)
   \label{mixing01},
   \end{equation}
  where ${\eta}_{q}$ $=$ $(u\bar{u}+d\bar{d})/{\sqrt{2}}$
  and ${\eta}_{s}$ $=$ $s\bar{s}$;
  the mixing angle ${\phi}$ ${\approx}$ $(39.3{\pm}1.0)^{\circ}$
  \cite{prd.58.114006}.
  Due to the symmetric flavor configurations of both ${\eta}_{q}$
  and ${\eta}_{s}$ states, we assume that DAs for ${\eta}_{q}$
  and ${\eta}_{s}$ states are similar to DAs for pion.
  It should be pointed out that the contributions from possible
  $c\bar{c}$ and gluonium compositions are not considered in our
  calculation for the moment, because
  (1) the final states with $B_{q}$ meson and $c\bar{c}$
  or gluonium states lie above the $B_{c}^{\ast}$ meson mass;
  (2) the fraction of gluonium components in ${\eta}$ and
  ${\eta}^{\prime}$ is rather tiny \cite{jhep0705.006}.
  Thus, the amplitudes for the $B_{c}^{\ast}$ ${\to}$ $B_{u}{\eta}$,
  $B_{u}{\eta}^{\prime}$ decays are written as
   \begin{eqnarray}
  {\cal A}(B_{c}^{\ast}{\to}B_{u}{\eta})
   &=&
   {\cos}{\phi}\,{\cal A}(B_{c}^{\ast}{\to}B_{u}{\eta}_{q})
  -{\sin}{\phi}\,{\cal A}(B_{c}^{\ast}{\to}B_{u}{\eta}_{s})
    \label{amp-bu-eta}, \\
   {\cal A}(B_{c}^{\ast}{\to}B_{u}{\eta}^{\prime})
   &=&
   {\sin}{\phi}\,{\cal A}(B_{c}^{\ast}{\to}B_{u}{\eta}_{q})
  +{\cos}{\phi}\,{\cal A}(B_{c}^{\ast}{\to}B_{u}{\eta}_{s})
    \label{amp-bu-etap}.
    \end{eqnarray}

  \subsection{Form factors}
  \label{sec0204}
  The hadron transition form factors are the basic input parameters for
  decay amplitudes [see Eq.(\ref{amp-bv-00}) and Eq.(\ref{amp-bp-00})].
  It is assumed \cite{npb591.313} that form factors come mainly from
  soft contributions, and form factors are generally
  regarded as nonperturbative parameters in the QCDF master formula
  of Eq.(\ref{hadronic}). Fortunately, form factors are universal.
  Form factors determined by other means or extracted from data
  can be employed here to make predictions.
  Phenomenologically, form factors are written as overlap
  integrals of wave functions.

  Here, we will employ the Wirbel-Stech-Bauer model \cite{zpc29.637}
  for evaluating the form factors.
  With a factorization of spin and spatial motion, wave function
  is written as
   \begin{equation}
  {\phi}^{(j,j_{z})}(\vec{k}_{\perp},x)\, =\,
  {\phi}(\vec{k}_{\perp},x)\, {\vert}s,s_{z};s_{1},s_{2}{\rangle}
   \label{bsw-phi-spin},
   \end{equation}
  where $\vec{k}_{\perp}$ and $x$ are the transverse momentum
  and longitudinal momentum fraction, respectively;
  $j$ ($s$) is the total angular momentum (spin);
  $j_{z}$ ($s_{z}$) is the magnetic quantum number;
  $s_{1}$ and $s_{2}$ are spins
  of valence quarks.
  $j$ $=$ $s$ $=$ $1$ for the ground vector $B_{c}^{\ast}$ meson,
  and $0$ for the ground pseudoscalar $B_{u,d,s}$ meson.
  The spatial wave function of a relativistic scalar harmonic
  oscillator potential is given by \cite{zpc29.637}
   \begin{equation}
  {\phi}(\vec{k}_{\perp},x)\, =\, N_{m}\, \sqrt{x\,(1-x)}\,
  {\exp}\Big\{-\frac{\vec{k}_{\perp}^{2}}{2\,{\omega}^{2}}\Big\}\,
  {\exp}\Big\{-\frac{m^{2}}{2\,{\omega}^{2}}\,\Big(x-\frac{1}{2}
  -\frac{m_{q_{1}}^{2}-m_{q_{2}}^{2}}{2\,m^{2}}\Big)^{2}\Big\}
   \label{bsw-phi},
   \end{equation}
  where parameter ${\omega}$ determines the average transverse momentum of
  partons, i.e., ${\langle}\vec{k}_{\perp}^{2}{\rangle}$ $=$ ${\omega}^{2}$;
  $m$ is the mass of the concerned meson; $m_{q_{1}}$ ($m_{q_{2}}$) is the
  constituent mass of the decaying (spectator) quark carrying a gluon cloud;
  $N_{m}$ is a normalization factor determined by
   \begin{equation}
  {\int}d^{2}k_{\perp}{\int}_{0}^{1}dx\,
  {\vert}{\phi}(\vec{k}_{\perp},x){\vert}^{2}\, =\, 1
   \label{bsw-nm}.
   \end{equation}

  \begin{figure}[ht]
  \includegraphics[width=0.98\textwidth,bb=75 610 530 710]{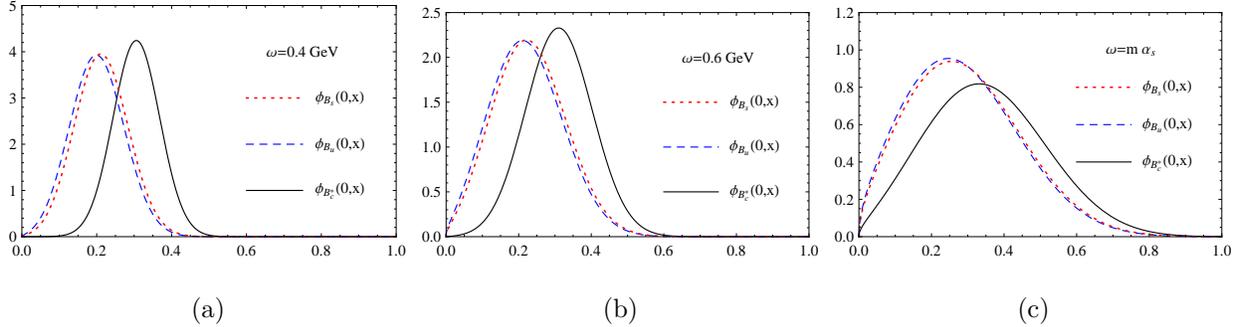}
  \caption{The shapes of the normalized wave functions
  for the $B_{c}^{\ast}$ and $B_{u,s}$ meson, with parameter ${\omega}$ $=$
  $0.4$ GeV (a), $0.6$ GeV (b), and $m\,{\alpha}_{s}$ (c),
  respectively.}
  \label{fig:wave}
  \end{figure}

  The form factors at zero momentum transfer are given by \cite{zpc29.637}
   \begin{equation}
  A_{0}(0)\, =\, A_{3}(0)\, =\,{\int}d^{2}k_{\perp}{\int}_{0}^{1}dx\,
  {\phi}_{B_{c}^{\ast}}^{(1,0)}(\vec{k}_{\perp},x)\,
  {\sigma}_{z}^{(1)}\,
  {\phi}_{B}(\vec{k}_{\perp},x)
   \label{bsw-ff-a0},
   \end{equation}
   \begin{equation}
  J\, =\, \sqrt{2}\,{\int}d^{2}k_{\perp}{\int}_{0}^{1}\frac{dx}{x}\,
  {\phi}_{B_{c}^{\ast}}^{(1,-1)}(\vec{k}_{\perp},x)\,
  i\,{\sigma}_{y}^{(1)}\,
  {\phi}_{B}(\vec{k}_{\perp},x)
   \label{bsw-ff-j},
   \end{equation}
   \begin{equation}
  V(0)\, =\, \frac{m_{c}-m_{q}}{ m_{B_{c}^{\ast}}-m_{B_{q}} }\,J
   \label{bsw-ff-v},
   \end{equation}
   \begin{equation}
  A_{1}(0)\, =\, \frac{m_{c}+m_{q}}{ m_{B_{c}^{\ast}}+m_{B_{q}} }\,J
   \label{bsw-ff-a1},
   \end{equation}
  where ${\sigma}_{z,y}^{(1)}$ are Pauli matrixes acting on the
  spin indices of the decaying quark $q_{1}$.

  It has been shown \cite{zpc29.637} that the form factors are
  sensitive to the choice of parameter ${\omega}$.
  And it is argued \cite{zpc29.637} that parameter ${\omega}$
  is not expected to be largely different for various mesons due
  to the flavor independence of the QCD interactions. Thus the same ${\omega}$
  might be applied to all mesons with the same spectator quark.
  The motion of the spectator (bottom) quark is nearly nonrelativistic
  in the $B_{c}^{\ast}$ ${\to}$ $B_{q}$ transition.
  Thus, nonrelativistic QCD (NRQCD) effective theory \cite{prd46.4052,prd51.1125,rmp77.1423}
  could be used to deal with both $B_{c}^{\ast}$ and $B_{q}$ mesons.
  According to the NRQCD power counting rules, the average transverse
  momentum is the order of ${\omega}$ ${\approx}$ $m\,{\alpha}_{s}$.
  In order to see the parameter ${\omega}$ effects on the form factors,
  we explore two scenarios. One is the same parameter ${\omega}$ for
  both the $B_{c}^{\ast}$ and $B_{q}$ mesons, and the other is ${\omega}$
  $=$ $m\,{\alpha}_{s}$, i.e., ${\omega}$ ${\approx}$ $1.24$ GeV for
  the $B_{c}^{\ast}$ meson, $1.10$ GeV for the $B_{s}$ meson, and
  $1.09$ GeV for the $B_{u,d}$ mesons.
  The numerical results for form factors are shown in Table \ref{tab:ff}.

  There are some comments on the form factors.
  (1) From the expressions in Eq.(\ref{bsw-ff-v}) and Eq.(\ref{bsw-ff-a1}),
  it is seen that due to the factor
  $\frac{m_{c}-m_{q}}{ m_{B_{c}^{\ast}}-m_{B_{q}} }$ ${\approx}$ $1$
  and $\frac{m_{c}+m_{q}}{ m_{B_{c}^{\ast}}+m_{B_{q}} }$ ${\ll}$ $1$,
  one can obtain a relation, $A_{1}(0)$ $<$ $V(0)$.
  (2)
  Compared with the integrand in Eq.(\ref{bsw-ff-a0}), there is a
  factor $1/x$ for the integrand in Eq.(\ref{bsw-ff-j}) with longitudinal
  momentum fraction $0$ $<$ $x$ $<$ $1$. Thus, it is expected to
  have in general $A_{0,3}(0)$ $<$ $V(0)$.
  (3)
  With the relation of form factors in Eq.(\ref{form01}),
  $A_{2}(0)$ is significantly enhanced by a factor of
  $\frac{2\,m_{B_{c}^{\ast}}}{ m_{B_{c}^{\ast}}-m_{B_{q}} }$
  (or $\frac{ m_{B_{c}^{\ast}}+m_{B_{q}} }{ m_{B_{c}^{\ast}}-m_{B_{q}} }$)
  relative to $A_{3}(0)$ (or $A_{1}(0)$).
  These relations are comprehensively verified by the numerical results
  for form factors in Table \ref{tab:ff}.

  In addition, from the numbers in Table \ref{tab:ff}, it is seen that
  (1) the form factors increase as parameter ${\omega}$ increases, due
  to the fact that the overlap between wave functions of $B_{c}^{\ast}$
  and $B_{q}$ mesons increases as parameter ${\omega}$ increases,
  as shown in Fig.\ref{fig:wave}.
  (2) The flavor symmetry breaking effects
  on form factors are small, but the isospin symmetry is basically held.
  (3) The values for $A_{2}(0)$ ($V(0)$) are about ten (five)
  times as large as those for $A_{1}(0)$, as explained above.
  The large values for 
  $A_{2}$ and $V$ would enhance the contributions from the $D$-
  and $P$-wave amplitudes (see Eq.(\ref{amp-bv-b})
  and Eq.(\ref{amp-bv-c})).

   \begin{table}[ht]
   \caption{The numerical values for the form factors in the $B_{c}^{\ast}$
   ${\to}$ $B_{q}$ transition, where the uncertainties come from both
   $m_{c}$ and $m_{b}$.}
   \label{tab:ff}
   \begin{ruledtabular}
   \begin{tabular}{cccccc}
    transition & ${\omega}$ & $A_{0}(0)$ & $A_{1}(0)$ & $A_{2}(0)$ & $V(0)$ \\ \hline
 & $0.4$ GeV
 & $0.540^{+0.015}_{-0.015}$ & $0.291^{+0.002}_{-0.002}$ & $3.286^{+0.210}_{-0.209}$ & $1.953^{+0.038}_{-0.040}$ \\
   $B_{c}^{\ast}$ ${\to}$ $B_{u}$
 & $0.6$ GeV
 & $0.784^{+0.008}_{-0.008}$ & $0.429^{+0.011}_{-0.011}$ & $4.694^{+0.219}_{-0.222}$ & $2.877^{+0.111}_{-0.109}$ \\
 & $m\,{\alpha}_{s}$
 & $0.944^{+0.002}_{-0.002}$ & $0.539^{+0.017}_{-0.017}$ & $5.403^{+0.208}_{-0.213}$ & $3.613^{+0.160}_{-0.156}$ \\ \hline
 & $0.4$ GeV
 & $0.540^{+0.015}_{-0.015}$ & $0.291^{+0.002}_{-0.002}$ & $3.288^{+0.210}_{-0.209}$ & $1.954^{+0.038}_{-0.040}$ \\
   $B_{c}^{\ast}$ ${\to}$ $B_{d}$
 & $0.6$ GeV
 & $0.784^{+0.008}_{-0.008}$ & $0.429^{+0.011}_{-0.011}$ & $4.696^{+0.219}_{-0.222}$ & $2.878^{+0.111}_{-0.109}$ \\
 & $m\,{\alpha}_{s}$
 & $0.944^{+0.002}_{-0.002}$ & $0.539^{+0.017}_{-0.017}$ & $5.405^{+0.208}_{-0.213}$ & $3.614^{+0.160}_{-0.156}$ \\ \hline
 & $0.4$ GeV
 & $0.609^{+0.015}_{-0.015}$ & $0.361^{+0.003}_{-0.003}$ & $3.618^{+0.234}_{-0.234}$ & $1.867^{+0.059}_{-0.061}$ \\
   $B_{c}^{\ast}$ ${\to}$ $B_{s}$
 & $0.6$ GeV
 & $0.821^{+0.007}_{-0.007}$ & $0.494^{+0.012}_{-0.012}$ & $4.785^{+0.242}_{-0.247}$ & $2.554^{+0.122}_{-0.120}$ \\
 & $m\,{\alpha}_{s}$
 & $0.954^{+0.002}_{-0.002}$ & $0.598^{+0.018}_{-0.017}$ & $5.268^{+0.232}_{-0.238}$ & $3.097^{+0.163}_{-0.159}$
   \end{tabular}
   \end{ruledtabular}
   \end{table}

  \section{Numerical results and discussion}
  \label{sec03}
  In the rest frame of the $B_{c}^{\ast}$ meson, branching ratios
  are defined as
   \begin{equation}
  {\cal B}r(B_{c}^{\ast}{\to}BV)\, =\, \frac{1}{24{\pi}}\,
   \frac{p_{\rm cm}}{m_{B_{c}^{\ast}}^{2}{\Gamma}_{B_{c}^{\ast}}}\,
   \big\{ {\vert}H_{+}{\vert}^{2}+{\vert}H_{0}{\vert}^{2}
   +{\vert}H_{-}{\vert}^{2} \big\}
   \label{br01},
   \end{equation}
   \begin{equation}
  {\cal B}r(B_{c}^{\ast}{\to}BP)\, =\, \frac{1}{24{\pi}}\,
   \frac{p_{\rm cm}}{m_{B_{c}^{\ast}}^{2}{\Gamma}_{B_{c}^{\ast}}}\,
  {\vert}{\cal A}(B_{c}^{\ast}{\to}BP){\vert}^{2}
   \label{br02},
   \end{equation}
  where ${\Gamma}_{B_{c}^{\ast}}$ is the full width of
  the $B_{c}^{\ast}$ meson.

  Because the electromagnetic radiation process $B_{c}^{\ast}$
  ${\to}$ $B_{c}{\gamma}$ dominates the $B_{c}^{\ast}$ meson
  decay, to a good approximation, ${\Gamma}_{B_{c}^{\ast}}$
  ${\simeq}$ ${\Gamma}(B_{c}^{\ast}{\to}B_{c}{\gamma})$.
  However, there is still no experimental information about the partial
  width ${\Gamma}(B_{c}^{\ast}{\to}B_{c}{\gamma})$ now,
  because the photon from the $B_{c}^{\ast}$ ${\to}$ $B_{c}{\gamma}$
  process is too soft to be easily identified.
  The information on ${\Gamma}(B_{c}^{\ast}{\to}B_{c}{\gamma})$
  comes mainly from theoretical estimation on the magnetic dipole
  (M1) transition, i.e., \cite{epja52.90}
   \begin{equation}
  {\Gamma}(B_{c}^{\ast}{\to}B_{c}{\gamma})\ =\
   \frac{4}{3}\,{\alpha}_{\rm em}\, k_{\gamma}^{3}\, {\mu}^{2}_{h}
   \label{m1-width},
   \end{equation}
  where ${\alpha}_{\rm em}$ is the fine-structure constant of
  electromagnetic interaction;
  $k_{\gamma}$ is the photon momentum in the rest frame of
  initial state; ${\mu}_{h}$ is the M1 moment of $B_{c}^{\ast}$
  meson. There are plenty of theoretical predictions on
  ${\Gamma}(B_{c}^{\ast}{\to}B_{c}{\gamma})$, for example,
  the numbers in Tables 3 and 6 in Ref.\cite{epja52.90}.
  However, these estimations still suffer from large uncertainties
  due to our lack of a precise value for ${\mu}_{h}$.
  To give a quantitative evaluation, ${\Gamma}_{B_{c}^{\ast}}$
  $=$ $50$ eV will be fixed in our calculation for the moment.
  The value of $50$ eV seems reasonable since it is close
  to the value given by the potential model (PM) which
  produces good agreement with experiment for the measured
  $J/{\psi}$ ${\to}$ ${\eta}_{c}{\gamma}$ decay rate.
  The value for the charm quark magnetic moment ${\mu}_{c}$
  obtained from the charmonium  M1 decay width can now be used
  to predict the $B_{c}^{\ast}$ ${\to}$ $B_{c}{\gamma}$ decay
  width, with a very small $b$ quark magnetic moment ${\mu}_{b}$
  $=$ $-0.06\,{\mu}_{N}$ given  in Ref.\cite{epja52.90}.

  The numerical values for other input parameters are listed in
  Table \ref{tab:input}. Unless otherwise stated,
  their central values will be fixed as the default inputs.
  Our numerical results are presented in Table \ref{tab:br}.
  The following are some comments.

   \begin{table}[ht]
   \caption{The numerical values for input parameters.}
   \label{tab:input}
   \begin{ruledtabular}
   \begin{tabular}{lll}
   \multicolumn{3}{c}{ Wolfenstein parameters \cite{pdg} } \\ \hline
   \multicolumn{3}{c}{
    ${\lambda}$ $=$ $0.22506{\pm}0.00050$, \qquad
    $A$ $=$ $0.811{\pm}0.026$, \qquad
    $\bar{\rho}$ $=$ $0.124^{+0.019}_{-0.018}$, \qquad
    $\bar{\eta}$ $=$ $0.356{\pm}0.011$; }  \\ \hline
    \multicolumn{3}{c}{Mass of particles and QCD characteristic scale} \\ \hline
    $m_{B_{c}^{\ast}}$ $=$ $6332{\pm}9$ MeV\footnotemark[1] \cite{prd86.094510},
  & $m_{{\pi}^{+}}$ $=$ $139.57$ MeV \cite{pdg},
  & $m_{{\pi}^{0}}$ $=$ $134.98$ MeV \cite{pdg}, \\
    $m_{B_{u}}$ $=$ $5279.31{\pm}0.15$ MeV \cite{pdg},
  & $m_{K^{+}}$ $=$ $493.677{\pm}0.016$ MeV \cite{pdg},
  & $m_{K^{0}}$ $=$ $497.611{\pm}0.013$ MeV \cite{pdg}, \\
    $m_{B_{d}}$ $=$ $5279.62{\pm}0.15$ MeV \cite{pdg},
  & $m_{\eta}$ $=$ $547.862{\pm}0.017$ MeV \cite{pdg},
  & $m_{{\eta}^{\prime}}$ $=$ $957.78{\pm}0.06$ MeV \cite{pdg}, \\
    $m_{B_{s}}$ $=$ $5366.82{\pm}0.22$ MeV \cite{pdg},
  & $m_{K^{{\ast}+}}$ $=$ $891.66{\pm}0.26$ MeV \cite{pdg},
  & $m_{K^{{\ast}0}}$ $=$ $895.81{\pm}0.19$ MeV \cite{pdg}, \\
    $m_{\rho}$ $=$ $775.26{\pm}0.25$ MeV \cite{pdg},
  & $m_{\omega}$ $=$ $782.62{\pm}0.12$ MeV \cite{pdg},
  & $m_{\phi}$ $=$ $1019.461{\pm}0.019$ MeV \cite{pdg}, \\
    $m_{b}$ $=$ $4.18^{+0.04}_{-0.03}$ GeV \cite{pdg},
  & $m_{c}$ $=$ $1.27{\pm}0.03$ GeV \cite{pdg},
  & ${\Lambda}^{(5)}_{\rm QCD}$ $=$ $210{\pm}14$ MeV \cite{pdg},\\
    $m_{s}$ $=$ $0.51$ GeV \cite{uds},
  & $m_{u,d}$ $=$ $0.31$ GeV \cite{uds},
  & ${\Lambda}^{(4)}_{\rm QCD}$ $=$ $292{\pm}16$ MeV \cite{pdg}, \\ \hline
    \multicolumn{3}{c}{Decay constants} \\ \hline
    $f_{\pi}$ $=$ $130.2{\pm}1.7$ MeV \cite{pdg},
  & $f_{K}$ $=$ $155.6{\pm}0.4$ MeV \cite{pdg},
  & $f_{K^{\ast}}$ $=$ $220{\pm}5$ MeV \cite{jhep.0703.069}, \\
    $f_{\rho}$ $=$ $216{\pm}3$ MeV \cite{jhep.0703.069},
  & $f_{\omega}$ $=$ $187{\pm}5$ MeV \cite{jhep.0703.069},
  & $f_{\phi}$ $=$ $215{\pm}5$ MeV \cite{jhep.0703.069}, \\
    $f_{q}$ $=$ $(1.07{\pm}0.02)\,f_{\pi}$ \cite{prd.58.114006},
  & $f_{s}$ $=$ $(1.34{\pm}0.06)\,f_{\pi}$ \cite{prd.58.114006}, \\ \hline
    \multicolumn{3}{c}{Gegenbauer moments at the sacle of ${\mu}$ $=$ 1 GeV} \\ \hline
    $a_{1}^{\pi}$ $=$ $a_{1}^{{\eta}_{q}}$ $=$ $a_{1}^{{\eta}_{s}}$ $=$ $0$ \cite{jhep.0605.004},
  & $a_{1}^{K}$ $=$ $0.06{\pm}0.03$ \cite{jhep.0605.004},
  & $a_{1}^{\rho}$ $=$ $a_{1}^{\omega}$ $=$ $a_{1}^{\phi}$ $=$ $0$ \cite{jhep.0703.069}, \\
    $a_{2}^{\pi}$ $=$ $a_{2}^{{\eta}_{q}}$ $=$ $a_{2}^{{\eta}_{s}}$ $=$ $0.25{\pm}0.15$ \cite{jhep.0605.004},
  & $a_{2}^{K}$ $=$ $0.25{\pm}0.15$ \cite{jhep.0605.004},
  & $a_{2}^{\rho}$ $=$ $a_{2}^{\omega}$ $=$ $0.15{\pm}0.07$ \cite{jhep.0703.069}, \\
    $a_{1}^{K^{\ast}}$ $=$ $0.03{\pm}0.02$ \cite{jhep.0703.069},
  & $a_{2}^{K^{\ast}}$ $=$ $0.11{\pm}0.09$ \cite{jhep.0703.069},
  & $a_{2}^{\phi}$ $=$ $0.18{\pm}0.08$ \cite{jhep.0703.069}.
   \end{tabular}
   \end{ruledtabular}
   \footnotetext[1]{Other predictions of the $B_{c}^{\ast}$ meson mass with
   different models can be found in Table II of  Ref.\cite{prd93.074010}.}
   \end{table}

   \begin{table}[ht]
   \caption{Branching ratios for the $B_{c}^{\ast}$ ${\to}$ $B_{q}V$,
   $B_{q}P$ decays calculated with the scale of ${\mu}$ $=$ $m_{c}$,
   where the parameters in the ``CKM'' (``$a_{i}$'') column
   give the CKM factors (coefficients) of the decay amplitude;
   the uncertainties come from mass $m_{c}$ and $m_{b}$.}
   \label{tab:br}
   \begin{ruledtabular}
   \begin{tabular}{lccccccc}
     \multicolumn{1}{c}{final} & \multicolumn{2}{c}{parameters} &
   & \multicolumn{4}{c}{branching ratio} \\ \cline{2-3} \cline{5-8}
     \multicolumn{1}{c}{state} & CKM & $a_{i}$ & case & ${\omega}$ $=$ $0.4$ GeV
   & ${\omega}$ $=$ $0.6$ GeV & ${\omega}$ $=$ $m\,{\alpha}_{s}$ & unit \\ \hline
   $B_{s}^{0}{\rho}^{+}$
 & $V_{cs}^{\ast}V_{ud}$ ${\sim}$ ${\cal O}(1)$ & $a_{1}$ & 1-I
 & $6.29^{+0.02}_{-0.02}$
 & $11.67^{+0.36}_{-0.35}$
 & $16.77^{+0.72}_{-0.68}$
 & $10^{-7}$ \\
   $B_{s}^{0}{\pi}^{+}$
 & $V_{cs}^{\ast}V_{ud}$ ${\sim}$ ${\cal O}(1)$ & $a_{1}$ & 1-I
 & $4.00^{+0.21}_{-0.20}$
 & $7.26^{+0.15}_{-0.15}$
 & $9.82^{+0.06}_{-0.06}$
 & $10^{-7}$ \\
   $B_{s}^{0}K^{{\ast}+}$
 & $V_{cs}^{\ast}V_{us}$ ${\sim}$ ${\cal O}({\lambda})$ & $a_{1}$ & 1-II
 & $2.21^{+0.02}_{-0.03}$
 & $4.12^{+0.18}_{-0.17}$
 & $6.01^{+0.32}_{-0.30}$
 & $10^{-8}$ \\
   $B_{s}^{0}K^{+}$
 & $V_{cs}^{\ast}V_{us}$ ${\sim}$ ${\cal O}({\lambda})$ & $a_{1}$ & 1-II
 & $1.98^{+0.10}_{-0.10}$
 & $3.60^{+0.08}_{-0.08}$
 & $4.87^{+0.03}_{-0.03}$
 & $10^{-8}$ \\ \hline
   $B_{d}^{0}{\rho}^{+}$
 & $V_{cd}^{\ast}V_{ud}$ ${\sim}$ ${\cal O}({\lambda})$ & $a_{1}$ & 1-II
 & $3.51^{+0.05}_{-0.05}$
 & $7.51^{+0.17}_{-0.17}$
 & $11.47^{+0.44}_{-0.41}$
 & $10^{-8}$ \\
   $B_{d}^{0}{\pi}^{+}$
 & $V_{cd}^{\ast}V_{ud}$ ${\sim}$ ${\cal O}({\lambda})$ & $a_{1}$ & 1-II
 & $2.14^{+0.13}_{-0.13}$
 & $4.50^{+0.11}_{-0.11}$
 & $6.53^{+0.05}_{-0.05}$
 & $10^{-8}$ \\
   $B_{d}^{0}K^{{\ast}+}$
 & $V_{cd}^{\ast}V_{us}$ ${\sim}$ ${\cal O}({\lambda}^{2})$ & $a_{1}$ & 1-III
 & $1.45^{+0.01}_{-0.01}$
 & $3.13^{+0.11}_{-0.11}$
 & $4.84^{+0.24}_{-0.23}$
 & $10^{-9}$ \\
   $B_{d}^{0}K^{+}$
 & $V_{cd}^{\ast}V_{us}$ ${\sim}$ ${\cal O}({\lambda}^{2})$ & $a_{1}$ & 1-III
 & $1.15^{+0.07}_{-0.07}$
 & $2.41^{+0.06}_{-0.06}$
 & $3.50^{+0.02}_{-0.02}$
 & $10^{-9}$ \\ \hline
   $B_{u}^{+}\overline{K}^{{\ast}0}$
 & $V_{cs}^{\ast}V_{ud}$ ${\sim}$ ${\cal O}(1)$ & $a_{2}$ & 2-I
 & $5.59^{+0.03}_{-0.03}$
 & $12.08^{+0.39}_{-0.36}$
 & $18.80^{+0.85}_{-0.80}$
 & $10^{-8}$ \\
   $B_{u}^{+}\overline{K}^{0}$
 & $V_{cs}^{\ast}V_{ud}$ ${\sim}$ ${\cal O}(1)$ & $a_{2}$ & 2-I
 & $3.48^{+0.29}_{-0.27}$
 & $7.32^{+0.34}_{-0.32}$
 & $10.60^{+0.31}_{-0.30}$
 & $10^{-8}$ \\
   $B_{u}^{+}K^{{\ast}0}$
 & $V_{cd}^{\ast}V_{us}$ ${\sim}$ ${\cal O}({\lambda}^{2})$ & $a_{2}$ & 2-III
 & $1.59^{+0.01}_{-0.01}$
 & $3.44^{+0.11}_{-0.10}$
 & $5.34^{+0.24}_{-0.23}$
 & $10^{-10}$ \\
   $B_{u}^{+}K^{0}$
 & $V_{cd}^{\ast}V_{us}$ ${\sim}$ ${\cal O}({\lambda}^{2})$ & $a_{2}$ & 2-III
 & $9.82^{+0.81}_{-0.77}$
 & $20.67^{+0.94}_{-0.90}$
 & $29.95^{+0.85}_{-0.82}$
 & $10^{-11}$ \\
   $B_{u}^{+}{\rho}^{0}$
 & $V_{cd}^{\ast}V_{ud}$ ${\sim}$ ${\cal O}({\lambda})$ & $a_{2}$ & 2-II
 & $1.84^{+0.04}_{-0.03}$
 & $3.95^{+0.07}_{-0.07}$
 & $6.08^{+0.20}_{-0.19}$
 & $10^{-9}$ \\
   $B_{u}^{+}{\omega}$
 & $V_{cd}^{\ast}V_{ud}$ ${\sim}$ ${\cal O}({\lambda})$ & $a_{2}$ & 2-II
 & $1.36^{+0.02}_{-0.02}$
 & $2.93^{+0.06}_{-0.05}$
 & $4.51^{+0.15}_{-0.14}$
 & $10^{-9}$ \\
   $B_{u}^{+}{\phi}$
 & $V_{cs}^{\ast}V_{us}$ ${\sim}$ ${\cal O}({\lambda})$ & $a_{2}$ & 2-II
 & $1.36^{+0.01}_{-0.01}$
 & $2.95^{+0.13}_{-0.12}$
 & $4.64^{+0.26}_{-0.25}$
 & $10^{-9}$ \\
   $B_{u}^{+}{\pi}^{0}$
 & $V_{cd}^{\ast}V_{ud}$ ${\sim}$ ${\cal O}({\lambda})$ & $a_{2}$ & 2-II
 & $9.24^{+0.77}_{-0.72}$
 & $19.44^{+0.89}_{-0.85}$
 & $28.17^{+0.81}_{-0.78}$
 & $10^{-10}$ \\
   $B_{u}^{+}{\eta}$
 & $V_{cd}^{\ast}V_{ud}$, $V_{cs}^{\ast}V_{us}$ & $a_{2}$ & 2-II
 & $2.37^{+0.20}_{-0.19}$
 & $4.98^{+0.23}_{-0.22}$
 & $7.22^{+0.21}_{-0.20}$
 & $10^{-9}$ \\
   $B_{u}^{+}{\eta}^{\prime}$
 & $V_{cd}^{\ast}V_{ud}$, $V_{cs}^{\ast}V_{us}$ & $a_{2}$ &
 & $6.62^{+0.55}_{-0.52}$
 & $13.93^{+0.64}_{-0.61}$
 & $20.18^{+0.58}_{-0.56}$
 & $10^{-11}$
   \end{tabular}
   \end{ruledtabular}
   \end{table}

  (1)
  According to the relative sizes of coefficients $a_{1,2}$ and CKM factors,
  the $B_{c}^{\ast}$ ${\to}$ $B_{q}V$, $B_{q}P$ decays could be classified
  into six cases (see Table \ref{tab:br}).
  There is a clear hierarchical relation among branching ratios,
  i.e., ${\cal B}r(\text{case 1-I})$ ${\sim}$ ${\cal O}(10^{-7})$,
  ${\cal B}r(\text{case 1-II})$ ${\sim}$ ${\cal O}(10^{-8})$,
  ${\cal B}r(\text{case 1-III})$ ${\sim}$ ${\cal O}(10^{-9})$,
  and ${\cal B}r(\text{case 2-I})$ ${\sim}$ ${\cal O}(10^{-8})$,
  ${\cal B}r(\text{case 2-II})$ ${\sim}$ ${\cal O}(10^{-9})$,
  ${\cal B}r(\text{case 2-III})$ ${\sim}$ ${\cal O}(10^{-10})$.

  (2)
  Branching ratios for the $B_{c}^{\ast}$ ${\to}$ $B_{q}V$ decays are
  generally larger than those for the $B_{c}^{\ast}$ ${\to}$ $B_{q}P$
  decays with the same final $B_{q}$ meson, where $V$ and $P$ have
  the same quark components.
  There are two reasons for this. One is the decay constant relation
  $f_{V}$ $>$ $f_{P}$, and the other is three partial wave contributions
  to the $B_{c}^{\ast}$ ${\to}$ $B_{q}V$ decays rather than only the $P$-wave
  contributions to the $B_{c}^{\ast}$ ${\to}$ $B_{q}P$ decays.

  It should be pointed out that although the $P$- and $D$-wave amplitudes
  for the $B_{c}^{\ast}$ ${\to}$ $B_{q}V$ decays are enhanced by large values
  for the form factors $V$ and $A_{2}$, they are suppressed by a factor of
  $\frac{2\,m_{B_{c}^{\ast}}\,m_{V}}{(m_{B_{c}^{\ast}}+m_{B_{q}})^{2}}$
  relative to the $S$-wave amplitude, as discussed above.
  In addition, the $P$- and $D$-wave contributions to helicity amplitudes
  $H_{\pm}$ in Eq.(\ref{amp-bv-hpm}) and $H_{0}$ in Eq.(\ref{amp-bv-h0})
  are future suppressed respectively by factors of $\sqrt{y^{2}-1}$ and
  $(y^{2}-1)/y$ relative to the $S$-wave contribution.
  Take the $B_{c}^{\ast}$ ${\to}$ $B_{s}{\rho}$ decay for example,
  $\frac{2\,m_{B_{c}^{\ast}}\,m_{\rho}}{(m_{B_{c}^{\ast}}+m_{B_{s}})^{2}}$
  ${\approx}$ 7\%, $\sqrt{y^{2}-1}$ ${\approx}$ $0.7$ and
  $(y^{2}-1)/y$ ${\approx}$ $0.4$, resulting in the polarization fractions
  $f_{0}$ ${\approx}$ 60\%, $f_{+}$ ${\approx}$ 30\% and $f_{-}$ ${\approx}$ 10\%
  with $f_{0,+,-}$ ${\equiv}$
  $\frac{{\vert}H_{0,+,-}{\vert}^{2}}{{\vert}H_{0}{\vert}^{2}+{\vert}H_{+}{\vert}^{2}+{\vert}H_{-}{\vert}^{2}}$.

  (3)
  The branching ratios for the $B_{c}^{\ast}$ ${\to}$ $B_{s}{\rho}$,
  $B_{s}{\pi}$ decays can reach up to ${\cal O}(10^{-7})$.
  With the estimated production cross section of the $B_{c}^{\ast}$ meson ${\sim}$
  $30$ $nb$ at LHC\cite{prd72.114009}, it is expected to have more than
  $10^{10}$ $B_{c}^{\ast}$ mesons per ${\rm ab}^{-1}$ data 
  at LHC, corresponding to more than $10^{3}$ events of the $B_{c}^{\ast}$
  ${\to}$ $B_{s}{\rho}$, $B_{s}{\pi}$ decays.
  Therefore, even with the identification efficiency, the $B_{c}^{\ast}$
  ${\to}$ $B_{s}{\rho}$, $B_{s}{\pi}$ decays might be measurable in the
  future.

  (4)
  Branching ratios for the $B_{c}^{\ast}$ ${\to}$ $B_{q}V$, $B_{q}P$ decays are
  several orders of magnitude smaller, especially for the $a_{1}$ dominant decays,
  than those for the $B_{c}$ ${\to}$ $B_{q}P$, $B_{q}V$ decays \cite{ahep2015.104378}.
  This fact might imply that possible background from the $B_{c}^{\ast}$ ${\to}$
  $BV$, $BP$ decays could be safely neglected for an analysis of the $B_{c}$
  ${\to}$ $B_{q}P$, $B_{q}V$ decays, but not vice versa, i.e., one of main pollution
  for the $B_{c}^{\ast}$ ${\to}$ $B_{q}V$, $B_{q}P$ decays would likely come from
  the $B_{c}$ decays.

  (5)
  It is seen clearly that the numbers in Table \ref{tab:br} are very sensitive
  to the choice of the parameter ${\omega}$.
  In addition, with a different value for ${\Gamma}_{B_{c}^{\ast}}$, branching
  ratios in Table \ref{tab:br} should be multiplied by a factor of
  ${50\,{\rm eV}}/{{\Gamma}_{B_{c}^{\ast}}}$.
  Of course, many factors, such as the choice of scale ${\mu}$,
  higher order corrections to HME, $q^{2}$-dependence of form factors,
  final state interactions, etc., are not carefully considered in detail here,
  but have effects on the estimation and deserve more dedicated study
  in the future.

  \section{Summary}
  \label{sec04}
  With the running and upgrading of the LHC, there are certainly huge
  amounts of the $B_{c}^{\ast}$ mesons. This would
  provide us with a possibility of
  searching for the $B_{c}^{\ast}$ weak decays in the future.
  In this paper, the $B_{c}^{\ast}$ ${\to}$ $B_{q}V$, $B_{q}P$
  decays ($q$ $=$ $u$, $d$ and $s$), induced by the charm quark weak decay,
  are studied phenomenologically with the QCDF approach.
  The form factors for the $B_{c}^{\ast}$ ${\to}$ $B$ transitions are
  calculated using the Wirbel-Stech-Bauer model.
  The nonfactorizable contributions from the vertex radiative corrections
  are considered at the order of ${\alpha}_{s}$.
  It is found that (1) form factors and branching ratios are sensitive
  to models of wave functions; (2) the color-favored and CKM-allowed
  $B_{c}^{\ast}$ ${\to}$ $B_{s}{\rho}$, $B_{s}{\pi}$ decays have large branching ratios of
  ${\cal O}(10^{-7})$, and might be accessible in the future LHC
  experiments.

  \section*{Acknowledgments}
  The work is supported by the National Natural Science Foundation
  of China (Grant Nos. U1632109, 11547014 and 11475055).
  We thank the referees for their constructive suggestions,
  and Ms. Nan Li (HNU) for polishing this manuscript.

  

\begin{thebibliography}{99}
  \bibitem{prd86.094510}
          R. Dowdall {\em et al.} (HPQCD Collaboration), Phys. Rev. D 86, 094510 (2012).
  \bibitem{epja52.90}
          V. \v{S}imonis, Eur. Phys. J. A 52, 90 (2016), and references therein.
  \bibitem{zpc51}
          M. Lusignoli, M. Masetti, Z. Phys. C 51, 549 (1991).
  \bibitem{prd49}
          C. Chang, Y. Chen, Phys. Rev. D 49, 3399 (1994).
  \bibitem{usp38}
          S. Gershtein {\em et al.}, Phys. Usp. 38, 1 (1995).
  \bibitem{prd77.074013}
          J. Sun {\em et al.}, Phys. Rev. D 77, 074013 (2008).
  \bibitem{prd89.114019}
          J. Sun {\em et al.}, Phys. Rev. D 89, 114019 (2014).
  \bibitem{ahep2015.104378}
          J. Sun {\em et al.}, Advances in High Energy Physics 2015, 104378 (2015).
  \bibitem{qwg}
          N. Brambilla {\em et al}. (Quarkonium Working Group), arXiv:hep-ph/0412158.
  \bibitem{plb355}
          K. Kolodziej, A. Leike, R. R\"{u}ckl, Phys. Lett. B 355, 337 (1995).
  \bibitem{plb364}
          C. Chang {\em et al.}, Phys. Lett. B 364, 78 (1995).
  \bibitem{prd54.4344}
          C. Chang, Y. Chen, R. Oakes, Phys. Rev. D 54, 4344 (1996).
  \bibitem{epjc38.267}
          C. Chang, X. Wu, Eur. Phys. J. C 38, 267 (2004).
  \bibitem{prd72.114009}
          C. Chang {\em et al.}, Phys. Rev. D 72, 114009 (2005).
  \bibitem{pdg}
          C. Patrignani {\em et al.} (Particle Data Group), Chin. Phys. C 40, 100001 (2016).
  \bibitem{prl83.1914}
          M. Beneke {\em et al.}, Phys. Rev. Lett. 83, 1914 (1999).
  \bibitem{npb591.313}
          M. Beneke {\em et al.}, Nucl. Phys. B 591, 313 (2000).
  \bibitem{npb606.245}
          M. Beneke {\em et al.}, Nucl. Phys. B 606, 245 (2001).
  \bibitem{plb488.46}
          D. Du, D. Yang, G. Zhu, Phys. Lett. B 488, 46 (2000).
  \bibitem{plb509.263}
          D. Du, D. Yang, G. Zhu, Phys. Lett. B 509, 263 (2001).
  \bibitem{prd64.014036}
           D. Du, D. Yang, G. Zhu, Phys. Rev. D 64, 014036 (2001).
  \bibitem{npb774.64}
          M. Beneke, J. Rohrer, D. Yang, Nucl. Phys. B 774, 64 (2007).
  \bibitem{npb832.109}
          M. Beneke, T. Huber, X. Li, Nucl. Phys. B 832, 109 (2010).
  \bibitem{plb750.348}
          G. Bell {\em et al.}, Phys. Lett. B 750, 348 (2015).
  \bibitem{prl.111}
          R. Aaij {\em et al.} (LHCb Collaboration), Phys. Rev. Lett 111, 181801 (2013).
  \bibitem{9512380}
          G. Buchalla, A. Buras, M. Lautenbacher, Rev. Mod. Phys. 68, 1125, (1996).
  \bibitem{npb11.325}
          J. Bjorken, Nucl. Phys. B (Proc. Suppl.) 11, 325 (1989).
  \bibitem{plb73.418}
          N. Cabibbo, L. Maiani, Phys. Lett. B 73, 418 (1978).
  \bibitem{npb133.315}
          D. Fakirov, B. Stech, Nucl. Phys. B 133, 315 (1978).
  \bibitem{zpc29.637}
          M. Wirbel, B. Stech, M. Bauer, Z. Phys. C 29, 637 (1985).
  \bibitem{zpc34.103}
          M. Bauer, B. Stech, M. Wirbel, Z. Phys. C 34, 103 (1987).
  \bibitem{prd63.014006}
          C. Bauer, S. Fleming, M. Luke, Phys. Rev. D 63, 014006 (2000).
  \bibitem{prd63.114020}
          C. Bauer {\em et al.}, Phys. Rev. D 63, 114020 (2001).
  \bibitem{plb516.134}
          C. Bauer, I. Stewart, Phys. Lett. B 516, 134 (2001).
  \bibitem{prd65.054022}
          C. Bauer, D. Pirjol, I. Stewart, Phys. Rev. D 65, 054022 (2002).
  \bibitem{prd66.014017}
          C. Bauer, {\em et al.}, Phys. Rev. D 66, 014017 (2002).
  \bibitem{npb643.431}
          M. Beneke {\em et al.}, Nucl. Phys. B 643, 431 (2002).
  \bibitem{plb553.267}
          M. Beneke, T. Feldmann, Phys. Lett. B 553, 267 (2003).
  \bibitem{npb685.249}
          M. Beneke, T. Feldmann, Nucl. Phys. B 685, 249 (2004).
  \bibitem{pqcd1}
          H. Li, Phys. Rev. D 52, 3958 (1995).
  \bibitem{pqcd2}
          C. Chang,  H. Li, Phys. Rev. D 55, 5577 (1997).
  \bibitem{pqcd3}
          T. Yeh, H. Li, Phys. Rev. D 56, 1615 (1997).
  \bibitem{prd22}
          G. Lepage, S. Brodsky, Phys. Rev. D 22, 2157 (1980).
  \bibitem{jhep.0605.004}
          P. Ball, V. Braun, A. Lenz, JHEP, 0605, 004 (2006).
  \bibitem{jhep.0703.069}
          P. Ball, G. Jones, JHEP, 0703, 069 (2007).
  \bibitem{npb268.16}
          A. Buras, J. Gerard, R. R\"{u}ckl, Nucl. Phys. B 268, 16 (1986).
  \bibitem{plb252.690}
          R. Verma, A. Kamal, A. Czarnecki, Phys. Lett. B 252, 690 (1990).
  \bibitem{ijmpa14.937}
          K. Sharma, R. Verma, Int. J. Mod. Phys. A 14,  937 (1999).
  \bibitem{epjc55.607}
          Y. Wang {\em et al.}, Eur. Phys. J. C 55, 607 (2008).
  \bibitem{prd81.074021}
          H. Cheng, C. Chiang, Phys. Rev. D 81, 074021 (2010).
  \bibitem{prd39.3339}
          G. Valencia, Phys. Rev. D 39, 3339 (1989).
  \bibitem{prd45.193}
          G. Kramer, W. Palmer, Phys. Rev. D 45, 193 (1992).
  \bibitem{prd.58.114006}
          Th. Feldmann, P. Kroll, B. Stech, Phys. Rev. D 58, 114006 (1998).
  \bibitem{jhep0705.006}
          R. Escribano, J. Nadal, JHEP 0705, 006 (2007).
  \bibitem{prd46.4052}
          G. Lepage {\em et al.}, Phys. Rev. D 46, 4052 (1992).
  \bibitem{prd51.1125}
          G. Bodwin, E. Braaten, G. Lepage, Phys. Rev. D 51, 1125 (1995).
  \bibitem{rmp77.1423}
          N. Brambilla {\em et al.}, Rev. Mod. Phys. 77, 1423 (2005).
  \bibitem{uds}
          A. Kamal, Particle Physics, Springer, 2014, p. 297, p. 298.
  \bibitem{prd93.074010}
          M. G\a'{o}mez-Rocha, T. Hilger, A. Krassnigg, Phys. Rev. D 93, 074010 (2016).
  \end{thebibliography}
  \end{document}